\documentclass[conference]{IEEEtran}
\usepackage{graphicx}
\usepackage{lipsum}
\usepackage{dirtytalk}
\usepackage[utf8]{inputenc}
\usepackage{subfigure}
\usepackage{mathtools}
\usepackage[algo2e,linesnumbered]{algorithm2e} 
\usepackage{algorithmic}
\usepackage{algorithm}
\usepackage{dirtytalk}
\usepackage[utf8]{inputenc}
\usepackage[english]{babel}
\usepackage{amsthm}
\usepackage{amsmath}
\usepackage{lipsum}
\newcommand{\subparagraph}{}
\usepackage{titlesec}
\usepackage{bbding}

\usepackage[colorlinks]{hyperref}
\providecommand{\keywords}[1]{\textbf{\textit{Index terms--}} #1}

\theoremstyle{definition}

\makeatletter
\newcommand*{\rom}[1]{\expandafter\@slowromancap\romannumeral #1@}
\makeatother

\showoutput
\titlespacing\section{0pt}{6pt plus 2pt minus 2pt}{6pt plus 2pt minus 2pt}
\titlespacing\subsection{0pt}{4pt plus 2pt minus 2pt}{4pt plus 2pt minus 2pt}
\titlespacing\subsubsection{0pt}{2pt plus 2pt minus 2pt}{2pt plus 2pt minus 2pt}

\begin{document}
\long\def\/*#1*/{}
\setlength{\abovedisplayskip}{1pt}
\setlength{\belowdisplayskip}{1pt}

\title{\textit{IoT}-enabled Channel Selection Approach for \textit{WBAN}s}

  \author{\IEEEauthorblockN{Mohamad Ali\IEEEauthorrefmark{1}, Hassine Moungla\IEEEauthorrefmark{2}, Mohamed Younis\IEEEauthorrefmark{3}, Ahmed Mehaoua\IEEEauthorrefmark{1}}
\IEEEauthorblockA{\IEEEauthorrefmark{1}LIPADE, University of Paris Descartes, Sorbonne Paris Cit\'{e}, Paris, France}{\IEEEauthorrefmark{2}UMR 5157, CNRS, Institute Mines-Telecom, T\'{e}l\'{e}com SudParis, Nano-Innov CEA Saclay, France}\IEEEauthorblockA{\IEEEauthorrefmark{3}Department of Computer Science and Electrical Engineering, University of Maryland, Baltimore County, United States}Email: \{mohamad.ali; hassine.moungla; ahmed.mehaoua\}@parisdescartes.fr; younis@umbc.edu}

\maketitle
\begin{abstract}
Recent advances in microelectronics have enabled the realization of \underline{\textit{W}}ireless \underline{\textit{B}}ody \underline{\textit{A}}rea \underline{\textit{N}}etworks (\textit{WBAN}s). However, the massive growth in wireless devices and the push for interconnecting these devices to form an \underline{\textit{I}}nternet \underline{\textit{o}}f \underline{\textit{T}}hings (\textit{IoT}) can be challenging for \textit{WBAN}s; hence robust communication is necessary through careful medium access arbitration. In this paper, we propose a new protocol to enable \textit{WBAN} operation within an \textit{IoT}. Basically, we leverage the emerging \underline{\textit{B}}luetooth \underline{\textit{L}}ow \underline{\textit{E}}nergy technology (\textit{BLE}) and promote the integration of a \textit{BLE} transceiver and a \underline{\textit{C}}ognitive \underline{\textit{R}}adio module (\textit{CR}) within the \textit{WBAN} coordinator. Accordingly, a \textit{BLE} informs \textit{WBAN}s through announcements about the frequency channels that are being used in their vicinity. To mitigate interference, the superframe's active period is extended to involve not only a \underline{\textit{T}}ime \underline{\textit{D}}ivision \underline{\textit{M}}ultiple \underline{\textit{A}}ccess (\textit{TDMA}) frame, but also a \underline{\textit{F}}lexible \underline{\textit{C}}hannel \underline{\textit{S}}election (\textit{FCS}) and a \underline{\textit{F}}lexible \underline{\textit{B}}ackup \textit{TDMA} (\textit{FBTDMA}) frames. The \textit{WBAN} sensors that experience interference on the \textit{default channel} within the \textit{TDMA} frame will eventually switch to another \underline{\textit{I}}nterference \underline{\textit{M}}itigation \underline{\textit{C}}hannel (\textit{IMC}). With the help of \textit{CR}, an \textit{IMC} is selected for a \textit{WBAN} and each interfering sensor will be allocated a time-slot within the (\textit{FBTDMA}) frame to retransmit using such \textit{IMC}.
\end{abstract}
\keywords{\textbf{{\small{\textit{\textit{IoT}},} \small{Channel allocation,} \small{WBAN interference mitigation,} \small{Bluetooth low energy,}  \small{Cognitive radio} }}}
\section{Introduction}
An \textit{IoT} is a short-range wireless network of interconnected devices, e.g.,\textit{WBAN}s, \textit{Wi-Fi},  \textit{IEEE 802.15.4} (\textit{ZigBee}), \textit{RFIDs}, \textit{Tags}, \textit{Sensors}, \textit{PDAs}, \textit{Smartphones}, etc., that could sense, process and communicate information. Example applications of \textit{IoT} are smart homes, health monitoring, wearables, environment monitoring, transportation and industrial automation. Within an \textit{IoT}, various types of wireless networks are required to facilitate the exchange of application-dependant data among their heterogeneous wireless devices. However, such diversity could give rise to coexistence issues among these networks, a challenge that limits the large-scale deployment of the \textit{IoT}. Therefore, new protocols are required for communication compatibility among its heterogeneous devices.

Basically, the \textit{IEEE 802.15.6} standard \cite{key1011}, e.g., \textit{WBANs}, utilizes a narrower bandwidth than other wireless networks, e.g., \textit{IEEE 802.11}. However, the \textit{IEEE 802.11} based wireless devices may use multiple channels that cover the whole international license-free \textit{2.4 GHz} \underline{\textit{I}}ndustrial, \underline{\textit{S}}cientific and \underline{\textit{M}}edical Radio, denoted by \textit{ISM}, band, so there could be overlapping channel covering an \textit{IEEE 802.15.6} based network and thus create collisions between \textit{IEEE 802.15.6} and these devices. In addition, \textit{IEEE 802.11} based wireless devices may transmit at a high power level and thus relatively distant coexisting \textit{IEEE 802.15.6} devices may still suffer interference. Thus, the pervasive growth in wireless devices and the push for interconnecting them can be challenging for \textit{WBAN}s due to their simple and energy-constrained nature. Basically, a \textit{WBAN} may suffer interference not only because of the presence of other \textit{WBAN}s but also from wireless devices within the general \textit{IoT} simultaneously operating on the same channel. Thus, co-channel interference may arise due to the collisions amongst the concurrent transmissions made by sensors in different \textit{WBAN}s collocated in an \textit{IoT} and hence such potential interference can be detrimental to the operation of \textit{WBAN}s. Therefore, robust communication is necessary among the individual devices of the collocated networks in an \textit{IoT}. 

In this paper, we propose a protocol to enable \textit{WBAN} operation within an \textit{IoT} and leverage the emerging \textit{BLE} technology to facilitate interference detection and mitigation. Motivated by the reduced power consumption and low cost of \textit{BLE} devices, we integrate a \textit{BLE} transceiver and a \textit{CR} module within each \textit{WBAN}'s coordinator node, denoted by \textit{Crd}, where the role of \textit{BLE} is to inform the \textit{Crd} about the frequency channels that are being used in its vicinity. In addition, the superframe's active period is further extended to involve not only a \textit{TDMA} frame, but also a \textit{FCS} and \textit{FBTDMA} frames, for interference mitigation. When experiencing high interference, the \textit{WBAN's Crd} will be notified by the \textit{BLE} device to use the \textit{CR} module for selecting a different channel. When engaged, the \textit{CR} assigns a stable channel for interfering sensors that will be used later within the \textit{FBTDMA} frame for data transmission. The simulation results show that our proposed approach can efficiently improve the spectrum utilization and significantly lower the medium access collisions among the collocated wireless devices in the general \textit{IoT}.

The rest of the paper is organized as follows. Section \rom{2} sets our work apart from other approaches in the literature. Section \rom{3} summarizes the system model and provides a brief overview of the \textit{BLE} and the \textit{CR}. Section \rom{4} describes \textit{CSIM} in detail. Section \rom{5} presents the simulation results. Finally, the paper is concluded in Section \rom{6}.
\section{Related Work}
Avoidance and mitigation of channel interference have been extensively researched in the wireless communication literature. To the best of our knowledge, the published techniques in the realm of \textit{IoT} are very few and can be categorized as resource sharing and allocation, power control, scheduling techniques and medium access schemes. Example schemes that pursued the resource sharing and allocation include \cite{key1000}, \cite{key1004}, \cite{key1002}, \cite{key1010}. Bakshi et al., \cite{key1000} proposed a completely asynchronous and distributed solution for data communication across \textit{IoT}, called \textit{EMIT}. \textit{EMIT} avoids the high overhead and coordination costs of existing solutions through employing an interference-averaging strategy that allows users to share their resources simultaneously. Furthermore, \textit{EMIT} develops power-rate allocation strategies to guarantee low-delay high-reliability performance. Torabi et al., \cite{key1004} proposed a rapid-response and robust scheme to mitigate the effect of interfering systems, e.g., \textit{IEEE 802.11}, on \textit{WBAN} performance. They proposed dynamic frequency allocation method to mitigate bi-link interferences that affect either the \textit{WBAN's Crd} or \textit{WBAN} sensors and hence impose them to switch to the same frequency. Shigueta et al., \cite{key1002} presented a strategy for channel assignment in an \textit{IoT}. The proposed strategy uses opportunistic spectrum access via cognitive radio. The originality of this work resides in the use of traffic history to guide the channel allocation in a distributed manner. Ali et al., \cite{key1010} proposed a distributed scheme that avoids interference amongst coexisting \textit{WBAN}s through predictable channel hopping. Based on the Latin rectangle of the individual \textit{WBAN}, each sensor is allocated a backup time-slot and a channel to use if it experiences interference such that collisions among different transmissions of coexisting \textit{WBAN}s are minimized. 

Xiao et al., \cite{key1001} adopted the approach of power control and considered machine-to-machine, denoted by \textit{M2M}, communication for an \textit{IoT} network. The authors proposed a framework of full-duplex \textit{M2M} communication in which the energy transfer, i.e., surplus energy, from the receiver to the transmitter and the data transmission from the transmitter to the receiver take place at the same time over the same frequency. Furthermore, the authors established a stochastic game-based model to characterize the interaction between autonomous \textit{M2M} transmitter and receiver. Meanwhile, Chen et al., \cite{key1003} introduced a new area packet scheduling technique involving \textit{IEEE 802.15.6} and \textit{IEEE 802.11} devices. The developed packet scheduler is based on transmitting a common control signal known as the blank burst from \textit{MAC} layer. The control signal prevents the \textit{IEEE 802.15.6} devices to transmit for a certain period of time during which the \textit{IEEE 802.11} devices could transmit data packets.

A number of approaches pursued the medium access scheduling methodology include \cite{key1006},\cite{key1007},\cite{key1005} to mitigate interference among the \textit{IEEE 802.11} and \textit{IEEE 802.15.4} \cite{key1012}, i.e., \textit{ZigBee}, based devices. Wang et al., \cite{key1006} proposed a new technique, namely, the Acknowledgement, denoted by \textit{ACK}, with Interference Detection (\textit{ACK-ID}), that reduces the \textit{ACK} losses and consequently reduces \textit{ZigBee} packet retransmissions due to the presence of collocated \textit{IEEE 802.11} wireless networks. Basically, in \textit{ACK-ID}, a novel interference detection process is performed before the transmission of each \textit{ZigBee} \textit{ACK} packet in order to decide whether the channel is experiencing interference or not. Inoue et.al., \cite{key1007} proposed a novel distributed active channel reservation scheme for coexistence, called \textit{DACROS}, to solve the problem of \textit{WBAN} and \textit{IEEE 802.11} wireless networks coexistence. \textit{DACROS} uses the request-to-send and clear-to-send frames to reserve the channel for a superframe time of \textit{WBAN}. Along the whole beacon time, i.e., the whole superframe of the \textit{WBAN}, all \textit{IEEE 802.11} wireless devices remain silent and do not transmit to avoid collisions. Zhang et al., \cite{key1005} proposed cooperative carrier signaling, namely, \textit{CCS}, to harmonize the coexistence of \textit{ZigBee} \textit{WBAN}s with \textit{IEEE 802.11} wireless networks. \textit{CCS} allows \textit{ZigBee} \textit{WBAN}s to avoid \textit{IEEE 802.11} wireless network-caused collisions and employs a separate \textit{ZigBee} device to emit a busy tone signal concurrently with the \textit{ZigBee} data transmission. 

As pointed out, none of the predominant approaches can be directly applied to \textit{IoT} because they do not consider the heterogeneity of the individual networks forming an \textit{IoT} in their design. Motivated by the emergence of \textit{BLE} technology and compared to the previous predominant approaches for interference mitigation, our approach lowers the power and communication overheads introduced on the coordinator- and sensor-levels within each \textit{WBAN}.

Unlike prior work, in this paper, we propose a distributed protocol to enable \textit{WBAN} operation and interaction within an existing \textit{IoT}. We integrate a \textit{BLE} transceiver to inform the \textit{WBAN} about the frequency channels that are being used in its vicinity and a \textit{CR} module within the \textit{WBAN's} \textit{Crd}. Our approach relies on both \textit{BLE} transceiver and the \textit{CR} module for stable channel selection and allocation for interference mitigation. The \textit{CR} module, when engaged determines a set of usable channels for the \textit{Crd} to pick from. Each interfering sensor will then switch to the new channel to retransmit data to the \textit{Crd} in its allocated backup time-slot.
\section{System Model and Preliminaries}
\subsection{Bluetooth Low Energy}
Bluetooth Low Energy (\textit{BLE}) is one of the promising technologies for \textit{IoT} services because of its low energy consumption and cost. \textit{BLE} is a wireless technology used for transmitting data over short distances and broadcasting advertisements at a regular interval via radio waves. The \textit{BLE} advertisement is a one-way communication method. \textit{BLE} devices, e.g., iBeacons, that want to be discovered can periodically broadcast self-contained packets of data. These packets are collected by devices like smartphones, where they can be used for a variety of applications to trigger prompt actions. We envision that each collocated set (cluster) of wireless devices of such \textit{IoT} will have to include a \textit{BLE} transceiver that periodically broadcasts the channel that is being used by the \textit{IoT} devices in the vicinity. In fact, with the increased popularity of \textit{BLE}, it is conceivable that every \textit{IoT} device will be equipped with a \textit{BLE} transceiver to announce its services and frequency channel. Standard \textit{BLE} has a broadcast range of up to 100 meters, which makes \textit{BLE} broadcasts an effective means for mitigating interference between \textit{WBAN}s and other \textit{IoT} devices.
\subsection{System Model and Assumptions}
The \textit{IoT} environment consists of different wireless networks, each uses some set of common channels in the international license-free \textit{2.4 GHz} \textit{ISM} band. In addition, we assume that each network transmits using different levels of transmission power, bandwidth, data rates and modulation schemes. Meanwhile, \textit{WBAN}s are getting pervasive and thus form a building block for the ever-evolving future \textit{IoT}. We consider \textit{N} \textit{TDMA}-based \textit{WBAN}s that coexist within the general \textit{IoT}. Each \textit{WBAN} consists of a single \textit{Crd} and up to \textit{K} sensors, each transmits its data on a channel within the international license-free \textit{2.4 GHz} \textit{ISM} band \cite{key1011}. Basically, we assume all \textit{Crds} are equipped with richer energy supply than sensors and all sensors have access to all \textit{ZigBee} channels at any time. In addition, each \textit{Crd} is integrated with \textit{BLE} to enable effective coordination in channel assignment and to allow the interaction with the existing \textit{IoT} devices. Furthermore, each \textit{Crd} has a \textit{CR} module to decide the usability and the stability of a channel.
\section{Channel Selection Approach for Interference Mitigation - \textit{CSIM}}
A co-channel interference takes place if the simultaneous transmissions of sensors and the \textit{Crd} in a \textit{WBAN} collide with those of other \textit{IoT} coexisting devices. The potential for such a collision problem grows with the increase in the communication range and the density of sensors in the individual \textit{WBAN}s as well as the number of collocated \textit{IoT} devices. To address this problem, our approach assigns each \textit{WBAN} a \textit{default channel} and in case of interference it allows the individual sensors to switch to a different channel to be picked by the \textit{Crd} in consultation with the \textit{CR} module to mitigate the interference. The use of \textit{BLE} enables the \textit{Crd} to be aware of interference conditions faster and more efficiently. To achieve that, our approach extends the size of the superframe through the addition of flexible number of backup time-slots to lower the collision probability of transmissions. At the network setup time, each \textit{Crd} randomly picks a \textit{default channel} from the set of \textit{ZigBee} channels and informs all sensors within its \textit{WBAN} through a beacon to use that channel along the \textit{TDMA} frame of the superframe, as will be explained below. %\textit{Crd} reports the selected channel through another beacon broadcast in the \textit{FCS} frame, as will be explained below.
\subsection{Network Operation under \textit{CSIM}}
\textit{CSIM} depends on acknowledgements (\textit{Acks}) and time-outs to detect the collision at \textit{sensor-} and \textit{coordinator-} levels. In the \textit{TDMA} frame shown in \textbf{Fig. \ref{superframeicc}}, each sensor transmits its packet in its assigned time-slot to the \textit{Crd} using the \textit{default channel} and then sets a time-out timer. If it successfully receives an \textit{Ack} from its corresponding \textit{Crd}, it considers the transmission successful, and hence it sleeps until the \textit{TDMA} frame of the next superframe. However, if that sensor does not receive an \textit{Ack} during the time-out period, it assumes failed transmission due to interference. Basically, all sensors experienced interference within the \textit{TDMA} frame wait until the \textit{FCS} frame completes, and then each switches to the common interference mitigation channel. Afterwards, each sensor retransmits its packet in its allocated time-slot within the \textit{FBTDMA} frame to the \textit{Crd}. \textbf{Algorithm \ref{csim}} provides high level summary of \textit{CSIM}. \textbf{Table \ref{symbol}} shows notations and their corresponding meanings. 
\begin{table}
\centering
\caption{Notations and meanings}
\label{symbol}
\begin{tabular}{lll}
\noalign{\smallskip}\hline
\textbf{Notation}&\textbf{Meaning}\\
\hline\noalign{\smallskip}
\textit{$WBAN_i$}&$i^{th}$ \textit{WBAN}\\
$S_{i,j}$&$j^{th}$ sensor of $i^{th}$ \textit{WBAN}\\
$defaultChannel_i$&default channel of $i^{th}$ \textit{WBAN}\\
$stableChannel_i$&stable channel of $i^{th}$ \textit{WBAN}\\
$\textit{Crd}_i$&coordinator of $i^{th}$ \textit{WBAN}\\
$BLE_i$&bluetooth low power device of $i^{th}$ coordinator\\
$CR_i$&cognitive radio module of $i^{th}$ coordinator\\
$Pkt_{i,j}$&$j^{th}$ packet of $i^{th}$ sensor\\
$Ack_{i,j}$&$i^{th}$ acknowledgement transmitted to $j^{th}$ sensor\\
$TS_{i,j}$&$j^{th}$ time-slot of $i^{th}$ \textit{TDMA} frame\\
$IMTS_{i,j}$&$j^{th}$ time-slot of $i^{th}$ \textit{FBTDMA} frame\\
$LCH_{i}$&$i^{th}$ set of channels used by nearby \textit{IoT} devices\\
$LIS_{i}$&$i^{th}$ list of interfering sensors in $TDMA_i$\\
\textit{FCS} & \textit{Flexible Channel Selection}\\
\textit{FBTDMA} & \textit{Flexible Backup TDMA}\\
\hline\noalign{\smallskip}
\end{tabular}
\end{table}
\subsection{Channel Selection}
Along the \textit{TDMA} frame, each \textit{Crd}'s \textit{BLE} collects information based on broadcast announcements made by other nearby \textit{BLE} transceivers about the set of channels being used by wireless devices in the vicinity of a designated \textit{WBAN} (\{\textit{LCH}\}), and then reports this information to its associated \textit{CR}. The \textit{CR} uses the following sets of channels which are defined as follows:
\begin{itemize}
    \item \{\textit{\textbf{G}}\} : is a set of \textit{16} channels available in the international license-free \textit{2.4 GHz} \textit{ISM} band of \textit{ZigBee} standard.
    \item \{\textit{\textbf{LCH}}\} : is a set of channels that are being used in the vicinity of a designated \textit{WBAN}.
    \item \{\textit{\textbf{defaultChannel}}\} : is a singleton set that involves the \textit{default channel} that is being used by a designated \textit{WBAN}.
    \item \{\textit{\textbf{US}}\} : is a set that consists of the remaining \textit{ZigBee} channels that are not being used in the vicinity of a designated \textit{WBAN}, where $\{\textit{US}\} = \{\textit{G}\} - \{\textit{\{LCH\}} \cup \textit{\{defaultChannel\}}$\}.
\end{itemize}

In low or moderate conditions of interference, where there are some available channels, i.e., \{\textit{US}\} is not empty, or the size of the set \{\textit{LCH}\} is smaller than the size of the set \{\textit{G}\}, the \textit{Crd} will not exploit the service of the \textit{CR} when notified by the \textit{BLE} about a channel conflict; instead, the \textit{Crd} selects one available channel from \{\textit{US}\} for efficient data transmission. However, in high interference conditions, the set \{\textit{US}\} will be empty. Therefore, once notified by the \textit{BLE}, the \textit{Crd} can not select one available channel from \{\textit{US}\}, and hence the \textit{CR} should scan the set \{\textit{LCH}\} to eventually select the most stable channel to be used within the \textit{FBTDMA} frame for interference mitigation. Basically, the designated \textit{CR} looks for a usable channel from the set \{\textit{LCH}\}, if the first channel is not, then it starts sequentially sensing channels until a usable channel will be found. If it finds a usable channel and satisfies the stability condition, then it reports its index to the associated \textit{Crd} to be eventually used for interference mitigation \cite{key998}.
\subsection{Channel Stability}
Our approach relies on \textit{CR} to decide the usability and stability of a channel using the received noise power as an indicator (\textit{$Y_i$}) \cite{key1009}. \textit{$Y_i$} during time-slot \textit{i} is given by \textbf{Eq. \ref{eq1}}.
\begin{equation}\label{eq1}
 Y_i = \frac{1}{2u}\sum_{j=1}^{2u}n_j \times n_j
\end{equation}
Where, \textit{u} is the time-bandwidth product and \textit{$n_j$} is a Gaussian noise signal with zero mean and unit variance. The probability density function, denoted by \textit{f}, of \textit{$Y_i$} is given by \textbf{Eq. \ref{eq2}}.
\begin{equation}\label{eq2}
  fY_{i}(y) = \frac{U}{\Gamma(.)}ke^{-uy}
\end{equation}
Where, \textit{$\Gamma(.)$} is the gamma function, \textit{$k=y^{u-1}$} and \textit{$U=u^{u}$}. Based on \textit{$Y_{i}$}, the \textit{CR} decision criterion can be expressed as follows:
\begin{enumerate}
\item A channel \textit{$C_{i}$} is usable, if \textit{$Y_i$ $<$ $\lambda_1$}
\item \textit{$C_{i}$} requires power boost (\textit{usable}), if \textit{$\lambda_1$ $<$ $Y_i$ $<$ $\lambda_2$}. In this case, we can use the theorem of Shannon (1948) \cite{key1008} of the maximum transmission capacity (\textit{P}) given in \textit{bit/s} in \textbf{Eq. \ref{eq4}}
\item \textit{$C_i$} cannot be used in time-slot \textit{i} (\textit{unusable}), if \textit{$Y_i$ $>$ $\lambda_2$}, where \textit{$\lambda_1$} and \textit{$\lambda_2$} are thresholds depend on the receiver sensitivity and the channel model in use.
\end{enumerate}
\begin{equation}\label{eq4}
 P = Blog_2 (1 + SNR)
 \end{equation}
Thus, the range of \textit{$Y_i$} is divided into three regions, and is given by \textbf{Eq. \ref{eq5}}. 
\begin{equation}\label{eq5}
R_j = \{Y_i : \lambda_{j-1} \leq Y_i \leq \lambda_j\}, j = 1, 2, 3
 \end{equation}
Where \textit{$\lambda_0$} is equal to \textit{0} and \textit{$\lambda_3$} is equal to \textit{$\infty$}. We mean by, a stable channel, if the probability of channel quality can not be decreased before the end of the transmission on that channel. The probability to being in a stable state \textit{j}  is given by \textbf{Eq. \ref{eq6}}.
\begin{equation}\label{eq6}
\pi_j = Pr\{Y_i \in R_j\} = Pr\{\lambda_{j-1} \leq Y_i < \lambda_j\}, j = 1, 2, 3
\end{equation}
The integration is done between \textit{$\lambda_{j-1}$} and \textit{$\lambda_j$}. When the \textit{CR} is engaged, it looks for a usable and stable channel which is done in the steps below.

\textbf{\textit{Step 1:}} \textit{Crd} looks for \textit{n} usable channels. If the first channel is not, then the \textit{CR} starts sequentially sensing channels until a usable channel is found. If the \textit{CR} module finds a usable channel, then \textbf{\textit{Step 2}} is executed to test the stability of the selected channel. Otherwise, the \textit{CR} module informs \textit{Crd} that no usable channel is available, \textit{Crd} stays silent during a predetermined time-slot.

\textbf{\textit{Step 2:}} If the selected usable channel satisfies the stability condition, then \textit{CR} reports the index of this stable channel back to \textit{Crd}.
\subsection{Proposed Superframe Structure}
\begin{figure}
  \centering
       \includegraphics[width=0.45\textwidth]{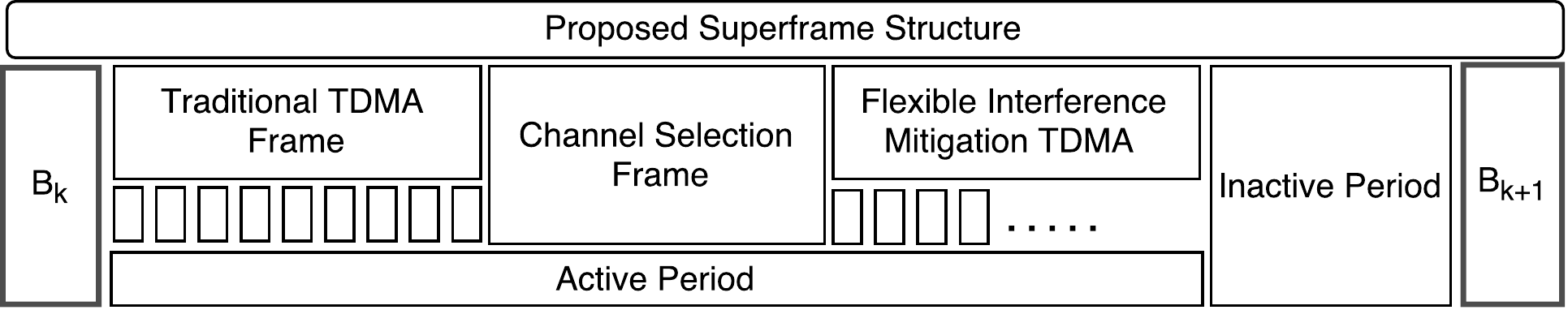}
\caption{Proposed superframe structure}
\label{superframeicc}
\end{figure}

In \textit{WBAN}s, sensors sleep and wake up dynamically and hence, the number of sensors being active during a period of time is unexpected. Therefore, a flexible way of scheduling different transmissions is required to avoid interference. We consider each \textit{WBAN}'s superframe delimited by two beacons and composed of two successive frames: (i) active, that is dedicated for sensors, and (ii) inactive, that is designated for \textit{Crd}s. The superframe structure is shown in \textbf{Fig. \ref{superframeicc}}. During the inactive frame, \textit{Crd}s transmit collected data to a command center. In addition, the inactive frame directly follows the active frame and whose length depends on the underlying duty cycle being used. However, the active frame is further divided into three successive frames.
\subsubsection{Traditional TDMA Data Collection Frame - \textit{TDMA}}
The traditional \textit{TDMA} frame consists of up to K time-slots that are allocated to sensors. Each \textit{WBAN}'s sensor transmits its packet to its associated \textit{Crd} in its allocated time-slot using the \textit{default channel}. 
\subsubsection{Channel Selection Frame - \textit{FCS}}
During the \textit{FCS} which is of a fixed size, each \textit{WBAN}'s \textit{Crd} selects a stable interference mitigation channel and instructs all interfering sensors within its \textit{WBAN} to use that channel during the \textit{FBTDMA} frame. Based on the number of interfering sensors, each \textit{Crd} determines the size of the \textit{FBTDMA} frame and reports this information through a short beacon broadcast using the \textit{default channel} to the designated sensors within its \textit{WBAN}. In addition, the \textit{Crd} allocates a time-slot within the \textit{FBTDMA} frame for each interfering sensor to eventually retransmit its packet. Although, the beacon could be lost due to the interference, our approach enables early mitigation. Basically, the \textit{BLE} alert limits the probability of collision on the \textit{default channel} since the \textit{Crd} will get a hint earlier than typical.
\subsubsection{Flexible Backup TDMA frame - \textit{FBTDMA}}
The \textit{FBTDMA} frame consists of a flexible number of backup time-slots that depends on the number of sensors experiencing interference in the \textit{TDMA} frame. Basically, each \textit{Crd} knows about these sensors through using the expected number of acknowledgement and data packets received in an allocated time-slot for each sensor. In \textit{FBTDMA} frame, each interfering sensor retransmits in its allocated backup time-slot to the \textit{Crd} using the selected stable channel. 
\begin{algorithm}
\caption{Proposed \textit{CSIM} Scheme}
\label{csim}
\begin{algorithmic}[1]
\REQUIRE{\textit{N WBAN}s, \textit{K Sensors/WBAN}, \textit{G \textit{ZIGBEE} Channels/WBAN}}         
          \STATE \textit{\textbf{Stage 1: Network Setup $\&$ TDMA Data Collection}} 
          \STATE \textbf{Sensor-level collision:} 
          \FOR{i = 1 to N}          
          \STATE \textit{$Crd_i$} picks one \textit{$defaultChannel_i$} from \textit{\{G\}}; 
          \FOR{J = 1 to K}  
           \STATE \textit{$S_{i,j}$} transmits $Pkt_{i,j}$ in \textit{$TS_{i,j}$} to \textit{$\textit{Crd}_i$} on \textit{$defaultChannel_i$};          
           \IF{\textit{$S_{i,j}$} receives \textit{$Ack_{i,j}$} on \textit{$defaultChannel_i$}}            
           \STATE \textit{$S_{i,j}$} sleeps until next superframe;            
           \ELSE            
           \STATE \textit{$S_{i,j}$} waits its \textit{$IMTS_{i,j}$} within \textit{$FBTDMA_{i}$} frame;           
           \ENDIF          
           \ENDFOR          
           \ENDFOR          
          \STATE \textbf{Coordinator-level collision:}          
          \FOR{i = 1 to N}             
          \FOR{j = 1 to K}             
          \IF {\textit{$Crd_i$} receives $Pkt_{i,j}$ in $TS_{i,j}$ on $defaultChannel_i$}            
          \STATE \textit{$Crd_i$} transmits $Ack_{i,j}$ in $TS_{i,j}$ to $S_{i,j}$ on $defaultChannel_i$;            
          \ELSE            
          \STATE \textit{$Crd_i$} will tune to $stableChannel_{i,j}$ within \textit{$FBTDMA_{i}$} frame;          
          \ENDIF          
          \ENDFOR          
          \ENDFOR          
          \STATE \textbf{Channel Selection Setup:}             
          \STATE $\textit{BLE}_i$ forms the set \textit{\{$LCH_i$\}};          
          \STATE $\textit{Crd}_i$ forms the set \textit{$\{LIS_i\}$};          
          \STATE \textit{\textbf{Stage 2: Channel Selection}}
          \FOR{i = 1 to N} 
          \STATE \textit{$Crd_i$} forms \textit{$FBTDMA_i$} frame from \textit{$\{LIS_i\}$};
          \STATE \textit{$CR_i$} selects $stableChannel_i$ from \textit{\{$US_i$\}};        
          \STATE \textit{$Crd_i$} informs \textit{$LIS_i$} sensors by \textit{$stableChannel_i$} \& \textit{$FBTDMA_i$} frame;         
          \ENDFOR         
          \STATE \textit{\textbf{Stage 3: Interference Mitigation}}
          \FOR{i = 1 to N} 
          \FOR{s = 1 to size-of($\{LIS_i\}$)}              
          \STATE $S_{i,s}$ retransmits $Pkt_{i,s}$ in $IMTS_{i,s}$ on $stableChannel_i$;            
           \IF{$Ack_{i,s}$ received by $S_{i,s}$ on $stableChannel_i$}          
          \STATE $S_{i,s}$ sleeps until next superframe;         
          \ELSE               
          \STATE $\textit{Crd}_i$ receives an earlier $\textit{BLE}_i$ alert of interference;
          \ENDIF
          \ENDFOR
          \ENDFOR
\end{algorithmic}
\label{csim}
\end{algorithm}
\section{Performance Evaluation}
In this section, we have conducted simulation experiments to evaluate the performance of the proposed \textit{CSIM} scheme. We compare the performance of \textit{CSIM} with smart spectrum allocation scheme \cite{key1012}, denoted by \textit{SSA}, which assigns orthogonal channels to sensors belonging to the interference set, denoted by \textit{IS}, formed between each pair of the interfering \textit{WBAN}s. Furthermore, we compare the energy consumption of the \textit{WBAN}'s coordinator with and without switching the \textit{BLE} transceiver on \cite{key1013}. We define the probability of channel's availability, denoted by \textit{$Pr_{AvChs}$}, at each \textit{Crd} as the frequency that a channel is not being used by any of the nearby \textit{IoT} devices. An \textit{IoT} cluster is defined as a collection of \textit{WBAN}s, \textit{Wi-Fi} and other wireless devices collocated in the same space. The simulation network is deployed in three dimensional space ($10\times10\times4 m^3$) and the locations of the individual \textit{WBAN}s change to mimic uniform random mobility and consequently, the interference pattern varies. The channel interference between any two wireless devices is evaluated on probabilistic interference thresholds. The simulation parameters are provided in \textbf{Table \ref{csimsp}}.
\begin{table}
\centering
\caption{Simulation parameters}
\label{csimsp}
\begin{tabular}{lllll}
\hline\noalign{\smallskip}
&\textbf{Exp. 1}&\textbf{Exp. 2}&\textbf{Exp. 3}\\
\hline\noalign{\smallskip}
\# Sensors/\textit{WBAN}&10&10&Var\\
\# \textit{WBAN}/network&Var&10&10\\
Sensor txPower (dBm)&-10&-10&-10\\
SNR threshold (dBm)&-25&Var&-25 \\
\# Time-slots/\textit{TDMA} frame& K&K&K\\
\noalign{\smallskip}\hline
\end{tabular}
\end{table}
\begin{figure*}
\begin{minipage}[b]{.3075\textwidth}
\centering
\includegraphics[width=1\textwidth, height=0.2\textheight]{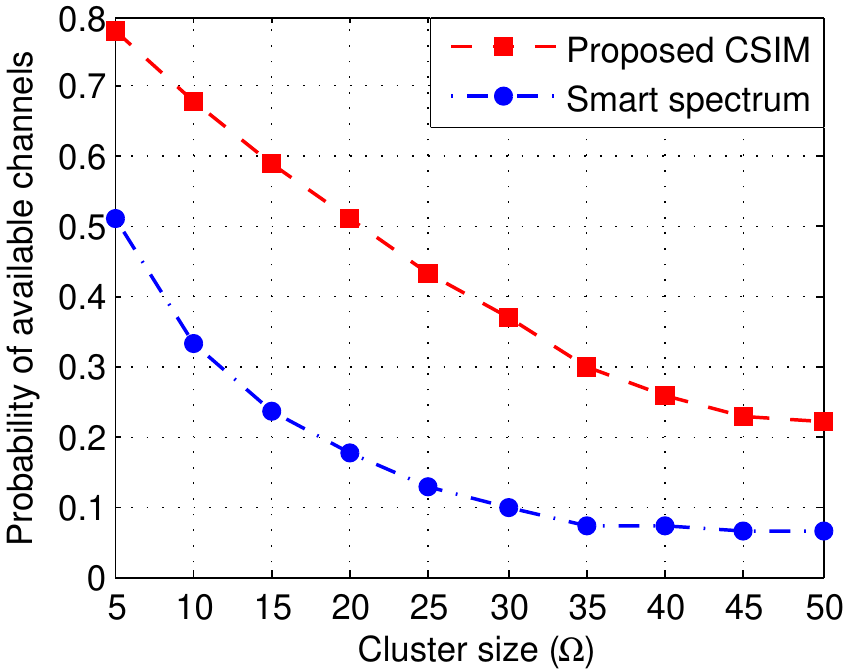}
\caption{Probability of channel's availability ($Pr_{AvChs}$) versus cluster size ($\Omega$)}
\label{plt1}
\end{minipage}\qquad
\begin{minipage}[b]{.3075\textwidth}
\centering
        \includegraphics[width=1\textwidth, height=0.2\textheight]{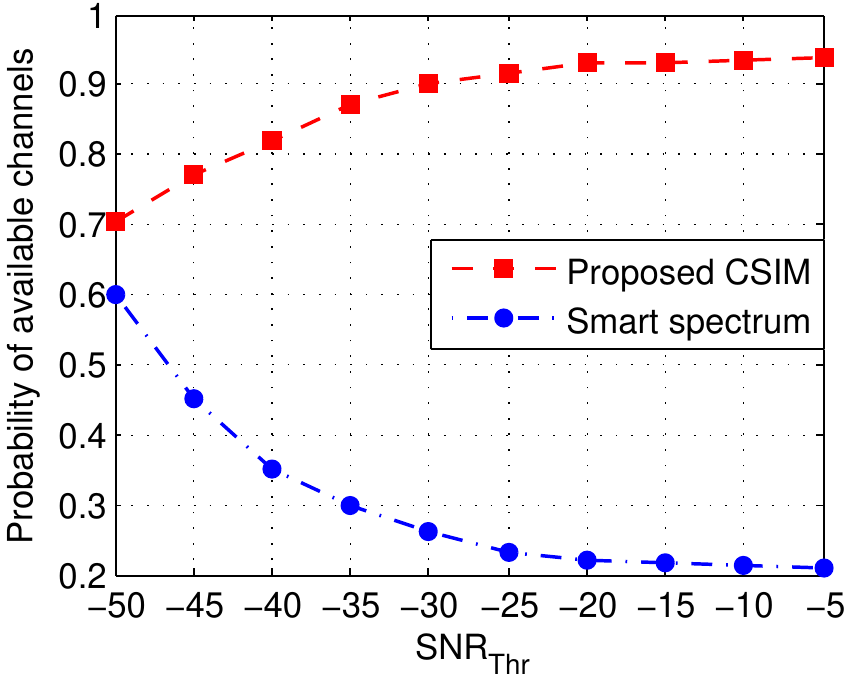}
\caption{$Pr_{AvChs}$ versus signal-to-noise ratio threshold ($SNR_{Thr}$)}
\label{plt2}
\end{minipage}\qquad
\begin{minipage}[b]{.3075\textwidth}
\centering
        \includegraphics[width=1\textwidth, height=0.2\textheight]{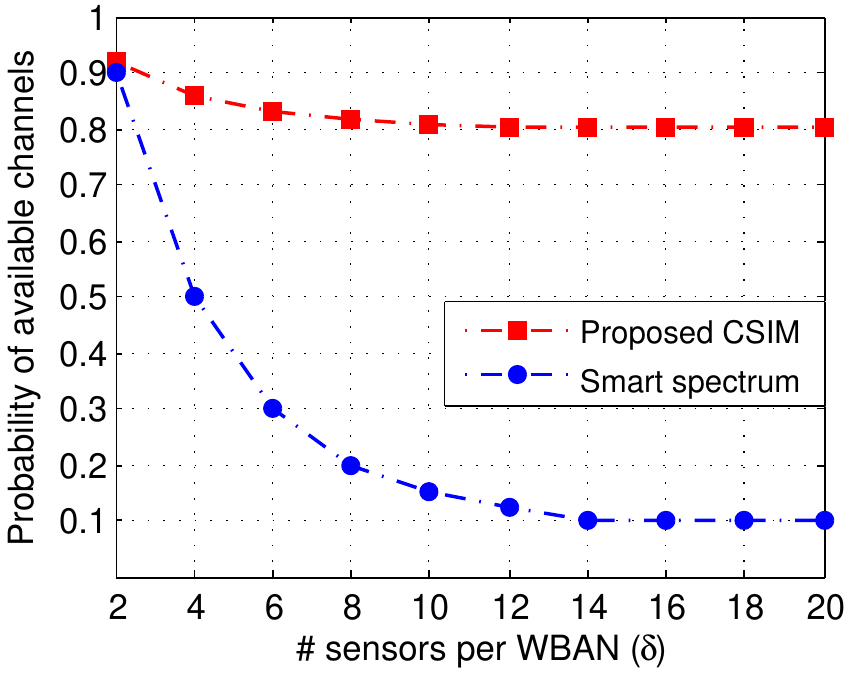}
\caption{$Pr_{AvChs}$ versus $\#$ of sensors per \textit{WBAN}($\delta$)}
\label{plt3}
\end{minipage}\qquad
\end{figure*}
\subsection{Probability of channel's availability}
\subsubsection{Probability of channel's availability vs. number of \textit{WBANs}}
In \textit{experiment 1}, the probability of channel's availability, denoted by \textit{$Pr_{AvChs}$}, versus the cluster size, denoted by $\Omega$, for \textit{CSIM} and \textit{SSA} are compared, and results are shown in \textbf{Fig. \ref{plt1}}. As seen in the figure, \textit{CSIM} always provides a higher \textit{$Pr_{AvChs}$} than \textit{SSA} because of the channel selection is done at the \textit{WBAN}- rather than \textit{sensor}-level. For \textit{CSIM}, the \textit{$Pr_{AvChs}$} significantly decreases from 0.79 to 0.27, when $5 \leq \Omega < 40$ because of the larger number of \textit{ZigBee} channels that are being used by \textit{IoT} devices than the number of channels available at each \textit{Crd}. When $\Omega \geq 40$, \textit{$Pr_{AvChs}$} decreases very slightly and eventually stabilizes at 0.215 because all \textit{ZigBee} channels are used by the \textit{IoT} devices which makes it very hard for \textit{Crd}s to select stable channels. However, for \textit{SSA}, it is also observed from this figure that \textit{$Pr_{AvChs}$} decreases significantly from 0.51 to 0.08 when $5 \leq \Omega < 35$ because of the larger number of \textit{ZigBee} channels that are being assigned to the sensors in the interfering set (\textit{IS}) for any pair of \textit{WBAN}s. When $\Omega \geq 35$, \textit{$Pr_{AvChs}$} decreases very slightly and eventually stabilizes at 0.07 because of the maximal number of \textit{ZigBee} channels being assigned to sensors coexisting within the interference range of a designated \textit{WBAN}, i.e., the number of these sensors exceeds the \textit{16} channels of \textit{ZigBee}.
\subsubsection{Probability of channel's availability vs. signal-to-noise ratio threshold}
\textit{Experiment 2} studies the effect of signal-to-noise ratio threshold denoted by \textit{$SNR_{Thr}$} on \textit{$Pr_{AvChs}$}. The results in \textbf{Fig. \ref{plt2}} shows that \textit{CSIM} always achieves higher \textit{$Pr_{AvChs}$} than \textit{SSA} for all \textit{$SNR_{Thr}$} values. In \textit{CSIM}, the \textit{$Pr_{AvChs}$} significantly increases as \textit{$SNR_{Thr}$} increases from $-50$ to $-35$; similarly increasing \textit{$SNR_{Thr}$} in \textit{CSIM} diminishes the interference range of each \textit{WBAN}, i.e., lowers the number of interfering \textit{IoT} devices. Therefore, limiting the frequency of channel assignments prevents distinct \textit{WBAN}s to pick the same channel, which decreases the probability of collisions among them. When \textit{$SNR_{Thr} \geq -35$}, the \textit{$Pr_{AvChs}$} increases very slightly and eventually stabilizes at 0.92 because of the minimal number of interfering \textit{IoT} devices and hence, a high \textit{$Pr_{AvChs}$} is expected due to the larger number of \textit{ZigBee} channels than the number of those interfering devices. However, \textit{SSA} always achieves lower \textit{$Pr_{AvChs}$} than \textit{CSIM} for all \textit{$SNR_{Thr}$} values. The \textit{$Pr_{AvChs}$} significantly decreases from 0.6 to 0.2 as \textit{$SNR_{Thr}$} increases from $-50$ to $-25$. Basically, increasing \textit{$SNR_{Thr}$} in \textit{SSA} is similar to increasing the interference range of each \textit{WBAN}, and hence putting more sensors in the \textit{WBAN} interference set. Therefore, more channels are needed to be assigned to those sensors and that \textit{$Pr_{AvChs}$} is reduced. When \textit{$SNR_{Thr} \geq -25$}, the \textit{$Pr_{AvChs}$} eventually stabilizes at 0.21 because of the maximal number of sensors in the interference set is attained by each \textit{WBAN}.
\subsubsection{Probability of channel's availability vs. number of sensors}
\textit{Experiment 3} studies the effect of the number ($\#$) of sensors per a \textit{WBAN}, denoted by \textit{$\delta$}, on \textit{$Pr_{AvChs}$}. As can be seen in \textbf{Fig. \ref{plt3}}, \textit{CSIM} always achieves higher \textit{$Pr_{AvChs}$} than \textit{SSA} for all values of $\delta$. It is also observed from this figure that \textit{$Pr_{AvChs}$} decreases very slightly and from 0.905 to 0.8 when $2 \leq \delta \leq 10$ and eventually stabilizes at 0.8 when $\delta \geq 10$. In both cases, the \textit{$Pr_{AvChs}$} is high due to two reasons, 1) the number of \textit{WBAN}s is fixed to \textit{10} which is smaller than the number of \textit{ZigBee} channels, which makes it possible for two or more distinct \textit{WBAN}s to not pick simultaneously the same channel and, 2) \textit{CSIM} selects a stable channel based on the number of interfering \textit{WBAN}s rather than the number of interfering sensors. However, the \textit{$Pr_{AvChs}$} decreases significantly from 0.9 to 0.1 when $2 \leq \delta \leq 14$ because adding more sensors into \textit{WBAN}s increases the probability of interference and consequently requires more channels to be assigned to those sensors; consequently \textit{$Pr_{AvChs}$} is reduced. Furthermore, \textit{SSA} assigns channels to interfering sensors rather than to interfering \textit{WBAN}s, which justifies the decrease of \textit{$Pr_{AvChs}$} when $\delta$ grows. When \textit{$\delta \geq 14$}, the \textit{$Pr_{AvChs}$} eventually stabilizes at 0.1 because of the maximal number of sensors in the interference set is attained by each \textit{WBAN}.
\begin{figure}
  \centering
       \includegraphics[width=0.3\textwidth, height=0.2\textheight]{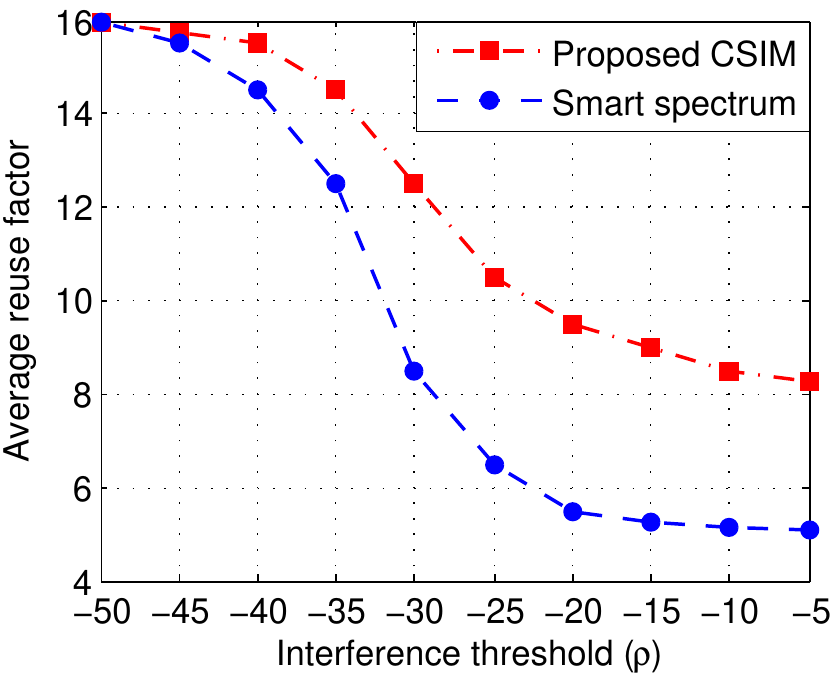}
\caption{Average reuse factor (\textit{avgRF} versus interference threshold ($\rho$)}
\label{plt4}
\end{figure}
\begin{figure}
  \centering
       \includegraphics[width=0.3\textwidth, height=0.2\textheight]{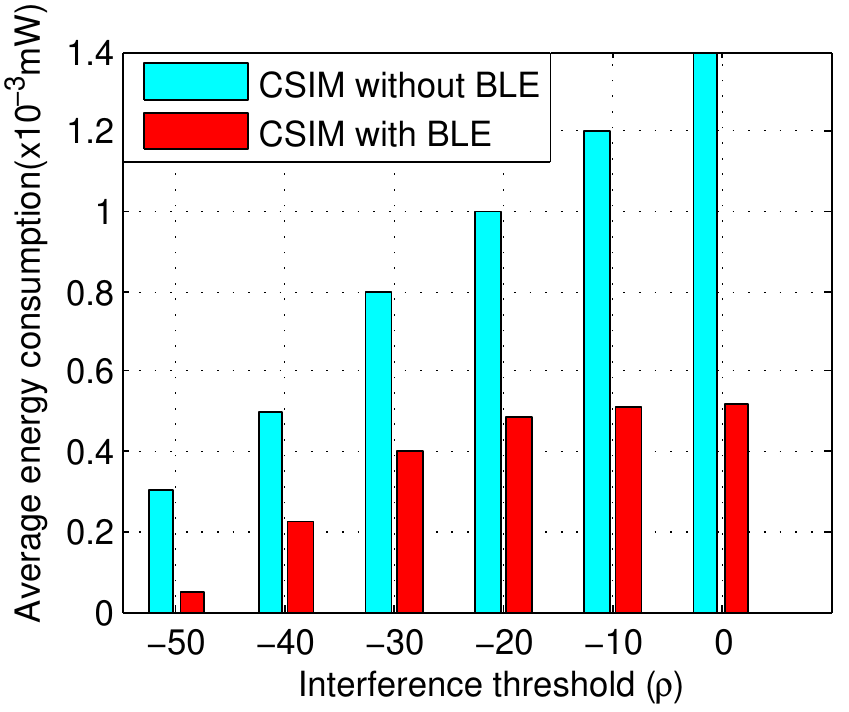}
\caption{Coordinator's average energy consumption (\textit{avgEC}) versus interference threshold ($\rho$)}
\label{plt5}
\end{figure}
\subsubsection{Average reuse factor vs. interference threshold}
\textbf{Fig. \ref{plt4}} shows the average reuse factor, denoted by \textit{avgRF}, versus the interference threshold, denoted by $\rho$, for all \textit{WBAN}s. As seen in this figure, \textit{CSIM} achieves a higher \textit{avgRF} for all $\rho$ values. However, increasing the interference threshold puts more interfering sensors in the interference range of any specific \textit{WBAN} than the corresponding \textit{WBAN}s of these sensors, i.e., \textit{SSA} requires more channels to be assigned to sensors than to \textit{WBAN}s in \textit{CSIM}.
\subsubsection{Energy consumption vs. interference threshold}
The average energy consumption of the \textit{WBAN} coordinator, denoted by \textit{avgEC}, versus the interference threshold ($\rho$) for \textit{CSIM} with (\textit{CSIM-W}) and without switching the \textit{BLE} transceiver \textit{on} (\textit{CSIM-WO}) are compared, and results are shown in \textbf{Fig. \ref{plt5}}. As seen in the figure, \textit{CSIM-W} always provides a lower \textit{avgEC} than \textit{CSIM-WO} because of the earlier \textit{BLE} alerts of interference to the coordinator, i.e., the coordinator scans the channels only upon receiving of these alerts. For \textit{CSIM-W}, the \textit{avgEC} increases slightly as the interference threshold grows, which increases the number of interfering sensors, hence the frequency of \textit{BLE} alerts of interference increases, and consequently, the energy consumption increases due to the additional scanning. When $\rho$ exceeds -20, the \textit{avgEC} increases very slightly and eventually stabilizes at $0.46\times10^{-3}$mW; this reflects the case where all channels are used by nearby \textit{IoT} devices forcing the \textit{Crd} to engage the \textit{CR} for finding a stable channel. For \textit{CSIM-WO}, the \textit{avgEC} increases significantly with all values of $\rho$ because of the continuous scanning of all \text{ZigBee} channels all the time, i.e., the coordinator periodically scans all the channels to find out which channels are not noisy. It is worth saying that the \textit{BLE} alerts reduces the frequency of channel scanning and hence saves the coordinator's energy.
\section{Conclusions}
In this paper, we have presented \textit{CSIM}, a distributed protocol to enable \textit{WBAN} operation and interaction within an existing \textit{IoT}. \textit{CSIM} leverages the emerging \textit{BLE} technology to enable channel selection and allocation for interference mitigation. In addition, the superframe’s active period is further extended to involve not only a \textit{TDMA} frame, but also a \textit{FCS} and \textit{FBTDMA} frames, for interference mitigation. We integrate a \textit{BLE} transceiver and a \textit{CR} within the \textit{WBAN}'s coordinator, where the role of the \textit{BLE} transceiver is to inform the \textit{WBAN} about the frequency channels that are being used in its vicinity. When experiencing high interference, the \textit{BLE} device notifies the \textit{WBAN's Crd} to call the \textit{CR} which determines a different channel for interfering sensors that will be used later within the \textit{FBTDMA} frame for interference mitigation. The simulation results show that \textit{CSIM} outperforms sample competing schemes.
 

\begin{thebibliography}{}

\end{thebibliography}


\begin{thebibliography}{16}
\bibitem{key1011}
IEEE Standard for Local and metropolitan area networks - Part 15.6: Wireless Body Area Networks: IEEE Std 802.15.6-2012
\bibitem{key1000}
Arjun Bakshi, Lu Chen, Kannan Srinivasan, Can Emre Koksal, Atilla Eryilmaz:
EMIT An efficient MAC paradigm for the Internet of Things. INFOCOM 2016: 1-9
\bibitem{key1004}
N. Torabi, W. K. Wong and V. C. M. Leung: A robust coexistence scheme for IEEE 802.15.4 wireless personal area networks. 2011 IEEE Consumer Communications and Networking Conference (CCNC), Las Vegas, NV, 2011, pp. 1031-1035.
\bibitem{key1002}
Roni F. Shigueta, Mauro Fonseca, Aline Carneiro Viana, Artur Ziviani, Anelise Munaretto:
A strategy for opportunistic cognitive channel allocation in wireless Internet of Things. Wireless Days 2014: 1-3
\bibitem{key1010}
Ali, M.J. and Moungla, H. and Younis, M. and Mehaoua, A.: Distributed Scheme for Interference Mitigation of WBANs Using Predictable Channel Hopping. 18th Int. Conf. on E-health Networking, Application \& Services (Healthcom): Munich, Germany. 2016
\bibitem{key1001}
Yong Xiao, Zixiang Xiong, Dusit Niyato, Zhu Han, Luiz A. DaSilva: Full-duplex machine-to-machine communication for wireless-powered Internet-of-Things. ICC 2016: 1-6
\bibitem{key1003}
Dong Chen, Jamil Y. Khan, Jason Brown:An area packet scheduler to mitigate coexistence issues in a WPAN/WLAN based heterogeneous network. ICT 2015: 319-325
\bibitem{key1006}
Zhipeng Wang, Tianyu Du, Yong Tang, Dimitrios Makrakis, Hussein T. Mouftah: ACK with Interference Detection Technique for ZigBee Network under Wi-Fi Interference. BWCCA 2013: 128-135
\bibitem{key1007}
Fumihiro Inoue, Masahiro Morikura, Takayuki Nishio, Koji Yamamoto, Fusao Nuno, Takatoshi Sugiyama: Novel coexistence scheme between wireless sensor network and wireless LAN for HEMS. SmartGridComm 2013: 271-276
\bibitem{key1005}
Xinyu Zhang, Kang G. Shin: Cooperative Carrier Signaling: Harmonizing Coexisting WPAN and WLAN Devices. IEEE/ACM Trans. Netw. 21(2): 426-439 (2013)
\bibitem{key1014}
IEEE Standard for Local and metropolitan area networks--Part 15.4: Low-Rate Wireless Personal Area Networks (LR-WPANs)," in IEEE Std 802.15.4-2011 (Revision of IEEE Std 802.15.4-2006) , vol., no., pp.1-314, Sept. 5 2011
\bibitem{key998}
Ala I. Al-Fuqaha, Mohsen Guizani, Mehdi Mohammadi, Mohammed Aledhari, Moussa Ayyash: Internet of Things: A Survey on Enabling Technologies, Protocols, and Applications. IEEE Communications Surveys and Tutorials 17(4): 2347-2376 (2015)
\bibitem{key1009}
Hassine Moungla, Kahina Haddadi, Saadi Boudjit: Distributed interference management in medical wireless sensor networks. CCNC 2016: 151-155
\bibitem{key1008},
Martial Coulon ENSEEIH:Systemes de telecommunications. 2007-2008
\bibitem{key1012}
Samaneh Movassaghi, Mehran Abolhasan, David B. Smith: Smart spectrum allocation for interference mitigation in Wireless Body Area Networks. ICC 2014: 5688-5693
\bibitem{key1013}
J. Lindhh, S. Kamath: Measuring Bluetooth Low Energy Power Consumption. Application Note AN092; Texas Instruments: Dallas, TX, USA, 2010

%\bibitem{key999}
%Gaoyang Shan, Sun-young Im, Byeong-Hee Roh:
%Optimal AdvInterval for BLE scanning in a different number of BLE devices environment. INFOCOM Workshops 2016: 1031-1032

\bibliographystyle{unsrt}
\end{thebibliography}
\end{document}